\documentclass{elsart}

\usepackage{graphicx}
\begin{document}
\begin{frontmatter}
%
%
\title{Correlated $\Lambda t$ pairs from the  absorption 
        of $K^-$ at rest in light nuclei.} 
\centering{\bf FINUDA Collaboration}
\author[polito,infnto]{M.~Agnello},
\author[infnba]{A.~Andronenkov},
\author[victoria]{G.~Beer},
\author[lnf]{L.~Benussi},
\author[lnf]{M.~Bertani},
\author[korea]{H.C.~Bhang},
\author[unibs,infnpv]{G.~Bonomi},
\author[unitos,infnto]{E.~Botta}, 
\author[units,infnts]{M. Bregant},
\author[unitos,infnto]{T.~Bressani},
\author[unitos,infnto]{S.~Bufalino},
\author[unitog,infnto]{L.~Busso},
\author[infnto]{D.~Calvo},
\author[units,infnts]{P.~Camerini},
\author[enea]{M. Caponero},
\author[uniba,infnba]{B.~Dalena},
\author[unitos,infnto]{F.~De~Mori},
\author[uniba,infnba]{G.~D'Erasmo}, 
\author[uniba]{D. Elia},
\author[lnf]{F.~L.~Fabbri},
\author[unitog,infnto]{D.~Faso}, 
\author[infnto]{A.~Feliciello}, 
\author[infnto]{A.~Filippi},  
\author[uniba,infnba]{M.~E.~Fiore}, 
\author[infnpv]{A.~Fontana},
\author[tokyo]{H.~Fujioka},
\author[lnf]{P.~Gianotti},  
\author[infnts]{N.~Grion}, 
\author[lnf]{O.~Hartmann}, 
\author[korea]{B.~Kang},
\author[jinr]{A.~Krasnoperov},
\author[korea]{Y.~Lee}, 
\author[uniba]{V. Lenti},
\author[lnf]{V.~Lucherini},
\author[infnba]{V. Manzari},  
\author[unitos,infnto]{S.~Marcello}, 
\author[tokyo]{T.~Maruta}, 
\author[teheran]{N.~Mirfakhrai},
\author[univpv,infnpv]{P.~Montagna}, 
\author[cnr,infnto]{O.~Morra}, 
\author[kyoto]{T.~Nagae}, 
\author[tokyo]{D.~Nakajima}
\author[riken]{H.~Outa}, 
\author[lnf]{E.~Pace}, 
\author[infnba]{M.~Palomba}, 
\author[infnba]{A.~Pantaleo}, 
\author[infnpv]{A.~Panzarasa}, 
\author[infnba]{V.~Paticchio}, 
\author[infnts]{S.~Piano\thanksref{corresponding}}, 
\author[lnf]{F.~Pompili},  
\author[units,infnts]{R.~Rui},
\author[kek]{M.~Sekimoto}, 
\author[uniba,infnba]{G.~Simonetti}, 
\author[jinr]{V. Tereshchenko},
\author[kek]{A. Toyoda},
\author[infnto]{R.~Wheadon}, 
\author[unibs]{A.~Zenoni}
\thanks[corresponding]{corresponding author. E-mail: stefano.piano@ts.infn.it; 
Fax:++39.040.5583350.}
\address[polito]{Dip. di Fisica Politecnico di Torino, Corso Duca degli 
Abruzzi, Torino, Italy}
\address[infnto]{INFN Sez. di Torino, via P.  Giuria 1, Torino, Italy}
\address[infnba]{INFN Sez. di Bari, via Amendola 179, Bari, Italy }
\address[victoria]{University of Victoria, Finnerty Rd.,Victoria, Canada}
\address[lnf]{Laboratori Nazionali di Frascati dell'INFN, via E. Fermi 40, 
Frascati, Italy}
\address[korea]{Dep. of Physics, Seoul National Univ., 151-742 Seoul, 
South Korea}
\address[unibs]{Dip. di Meccanica,  Universit\`a di Brescia, Via vallotti 9, 
Brescia, Italy}
\address[infnpv]{INFN Sez. di Pavia, Via Bassi 6, Pavia, Italy}
\address[unitos]{Dipartimento di Fisica Sperimentale, Universit\`a di
Torino, via P. Giuria 1 Torino, Italy} 
\address[units]{Dip. di Fisica Univ. di Trieste, via Valerio 2, Trieste, Italy}
\address[infnts]{INFN Sez. di Trieste,  via Valerio 2, Trieste, Italy}
\address[unitog]{Dipartimento di Fisica Generale, Universit\`a di Torino, 
via P. Giuria 1, Torino, Italy} 
\address[enea]{ENEA, Frascati, Italy}
\address[uniba]{Dip. di Fisica Univ. di Bari, via Amendola 179 Bari, Italy }
\address[tokyo]{Dep. of Physics Univ. of Tokyo, Bunkyo Tokyo 113-0033, Japan}
\address[jinr]{Joint Institute for Nuclear Research (JINR), Dubna, Russia}
\address[teheran]{Dep of Physics Shahid Behesty Univ., 19834 Teheran, Iran}
\address[univpv]{Dipartimento di Fisica Teorica e Nucleare, Universit\a`
di Pavia, Via Bassi 6, Pavia, Italy}
\address[cnr]{INAF-IFSI Sez. di Torino, C.so Fiume 4, Torino, Italy}
\address[kyoto]{Dep. of Physics Sakyo-ku, Kyoto 606-8502, Japan}
\address[riken]{RIKEN, Wako, Saitama 351-0198, Japan}
\address[kek]{High Energy Accelerator Research Organization (KEK), Tsukuba, 
Ibaraki 305-0801 Japan}
%
%
\begin{abstract}
Novel data from the $K^{-}_{stop}A$ absorption reaction in light nuclei 
$^{6,7}$Li and  $^{9}$Be are presented.  The study aimed at finding 
$\Lambda t$ correlations. Regardless of $A$, the $\Lambda t$ pairs are 
preferentially emitted in opposite directions. Reaction modeling 
predominantly assigns to the $K^-_{stop}A\rightarrow\Lambda t(N)A'$ 
direct reactions the emission of the $\Lambda t$ pairs whose yield is 
found to range from $10^{-3}$ to $10^{-4}$$/K^-_{stop}$. The experiment 
was performed with the FINUDA spectrometer at DA$\Phi$NE (LNF).
\bigskip

{\it PACS:21.45.+v, 21.80.+a 25.80.Nv}
\end{abstract}
\end{frontmatter}
%
%
\section{Introduction}
In this Letter we present the results of an analysis of correlated 
$\Lambda$-hyperon and triton ($\Lambda t$) pairs following the 
$K^{-}_{stop}A$ absorption reaction on several nuclei. The data were 
collected by the FINUDA spectrometer running at the Laboratori Nazionali 
di Frascati (LNF), Italy. Data on reactions of $K^-$ nuclear absorption 
with the emission of nucleons and nuclei are scarce. The bulk of the 
available data belongs to bubble chamber experiments, which mainly aimed 
at assessing the $K^-$ capture rates of multipionic final states 
\cite{expt:katz}. A handful of measurements studied the $K^-$-absorption 
reaction leading to non-mesonic channels \cite{expt:roosen}. This 
investigation received a strong boost recently, when non-mesonic 
channels were related to the non-mesonic decay mode of 
$\overline K$-nuclear bound (KNB) states. The existence of such states 
is presently being debated within the hadron physics community. 
According to the theoretical models, KNB states preferentially decay to 
a hyperon and one ($N$) or more nucleons ($nN$), the decay into pionic 
channels being suppressed by the strong binding energy of such states 
\cite{theor:akaishi,theor:gadza}. Therefore, the invariant mass studies 
of $\Lambda N(nN)$ pairs could provide direct evidence of their 
existence. Inclusive or semi-inclusive spectra such as the energy 
spectrum of $N$ also supply information about the nature of KNB states; 
however, the sizable continuous background and the meager content of 
information may lead to erroneous interpretation of results. As earlier 
pointed out, the K-absorption reaction on nucleon clusters leading to 
non-mesonic final states is still poorly examined. Therefore, further  
experimental studies are needed, which may clarify the mechanism of
kaon absorption on multibarionic systems. 

Experimental studies on the existence of $K^-$-nuclear bound states 
were performed by the FINUDA collaboration by examining the invariant 
mass distribution of $\Lambda N(2N)$ pairs 
\cite{expt:FINUDA0,expt:FINUDA2}. These distributions show bumps 
standing over a continuous background. The width of these bumps is 
several tens of MeV, and their strength lies below the sum of the 
$K^-NN(NNN)$ rest masses. Moreover, the $\Lambda$-hyperon was found to 
be strongly back-to-back correlated with the selected nucleon(s), which 
indicates that the negative kaon strongly couples to the $NN(NNN)$ 
nucleon clusters before decaying to $\Lambda N(2N)$ pairs. All these 
findings are compatible with the formation of bound $[K^-NN(NNN)]$ 
clusters. A full kinematic analysis of the $K^-$-induced reactions in 
nuclei is difficult because of the limited experimental information 
about the residual nuclear fragments. A study of $\Lambda d$ pairs 
emitted after $K^-$ absorption at rest on $^4$He was recently performed 
by the KEK-PS E549 collaboration \cite{expt:suzuki}. The observed 
invariant mass distribution of the back-to-back correlated $\Lambda d$ 
pairs displays features which resemble those of the FINUDA invariant 
mass distribution for $^6Li$; for instance, compare Fig. 3 of Refs. 
\cite{expt:suzuki} and \cite{expt:FINUDA2}, respectively. 

Following this trail, a study of $\Lambda t$ coincidence events from 
the $K^-_{stop}A\rightarrow\Lambda t A'$ reaction was pursued by FINUDA 
and will be presented in this Letter along with a discussion on several 
reaction mechanisms which can explain the observed data. To our knowledge, 
only one old measurement of $K^-$ at rest induced interactions leading to 
final $\Lambda t$ pairs exists \cite{expt:roosen}. In this measurement, 
only three events (out of 3258) were found to be compatible with the 
$K^-_{stop}$$^4He\rightarrow\Lambda t$ reaction. 
\section{The Experimental Method}
In this study, the $\Lambda t$ events from the absorption reaction 
$K^-_{stop}A\rightarrow\Lambda t A'$ are reconstructed in FINUDA.  The 
residual nucleus, $A'$, is regarded as a system of $[A-4]$ nucleons 
whose final state is not reconstructed. Due to the poor $\Lambda t$ 
statistics, the contribution of all the targets, $^{6,7}$Li and 
$^{9}$Be, is summed.  

The experimental method is briefly explained in this Letter; further 
details are reported in Refs. \cite{expt:FINUDA2,FINUDA:01}. Negative 
(positive) kaons of (16.1$\pm$1.5) MeV come from the process 
$e^+e^-\rightarrow\phi(1020)\rightarrow K^+K^-$ (B.R.$\sim$50\%), where 
the $\phi$ mesons are created by 510 MeV electron-positron collisions 
at the DA$\Phi$NE collider at LNF. The $K^-$'s slow down as they cross 
some of the inner layers of the magnetic spectrometer until they stop 
within solid targets, which are as thin as 0.2-0.3 g/cm$^2$. The magnetic 
field was set at 1.0 T. The spectrometer has cylindrical geometry around 
the $e^+e^-$ axes. It consists of several sensitive layers which are 
used either for particle identification (d$E$/d$x$), or for particle 
localization, or for both. The first layer encountered by the kaons is 
TOFINO \cite{FINUDA:02}. It is a segmented detector made of plastic 
scintillator, which is optimized for starting the time-of-flight system 
and for trigger purposes. TOFINO is followed by ISIM and OSIM 
\cite{FINUDA:03}, two layers of double-sided silicon strip detectors, 
which are used  for both localization and identification of charged 
particles. The eight nuclear targets are located between ISIM and OSIM 
which are surrounded by two layers of low-mass drift chambers (LMDC), 
which localize charged particles and contribute to their identification 
\cite{FINUDA:04}. A system of six stereo-arranged layers of straw tubes 
(ST) is the last position sensitive tracking device located within the 
magnetic field \cite{FINUDA:05}. The outermost  layer of FINUDA, TOFONE, 
consists of 72 trapezoidal slabs of plastic scintillator \cite{FINUDA:06}. 
TOFONE is the stop-counter of the FINUDA time-of-flight measurements and 
also measures the energy released by charged particles and neutrons. 

%
%
\begin{figure}[b]
 \centering
  \includegraphics*[angle=0,width=0.8\textwidth]
   {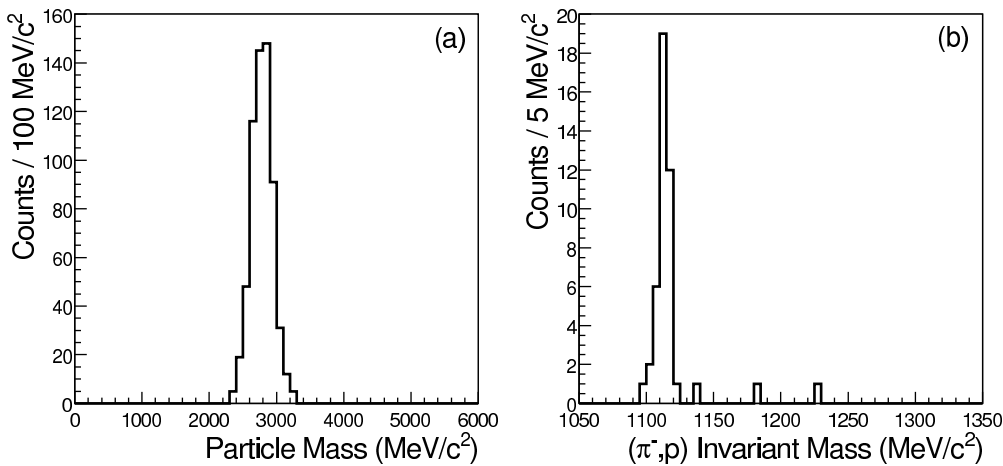}
    \caption{(a) Triton mass identification.(b) Invariant mass 
      distribution of $\pi^- p$ pairs in coincidence with tritons.   
      Details are given in the text.}
\end{figure}   
The mass identification (ID) of $\pi^- $'s, $p$'s, $d$'s and $t$'s relies  
on the specific energy loss (d$E$/d$x$) of these particles in some of the 
spectrometer layers; namely, I/OSIM, LMDC's and TOFONE. Because of the 
high specific energy loss of tritons, (d$E$/d$x)_t$, their trajectories 
are reconstructed in about half of the FINUDA solid angle, which comprises 
OSIM, two layers of LMDC's and TOFONE (forward tritons). Backward tritons, 
before being tracked, must cross the material budget of the FINUDA central 
region where they nearly come to a stop. Forward tritons are initially 
selected according to their specific energy loss in OSIM and LMDC's: 
(d$E$/d$x)_t>$(d$E/dx)_d$. This approach allows minimum ionizing particles 
and (a large fraction of) protons and deuterons to be discarded, which are 
the overwhelming number of particles in this measurement. In addition, a 
triton impinges on TOFONE when its momentum ($p_t$) exceeds 657 MeV/$c$. 
For this triton, a valid signal from TOFONE is then requested. Below this 
momentum, tritons do not have an effect on TOFONE, which provides a method 
to mass-identify tritons of lower momentum: they can be identified by a 
valid (d$E$/d$x)_t$ response  not followed by a signal from TOFONE. This 
approach lowers the $p_t$ threshold to $\sim$430 MeV/$c$ since tritons 
are required to cross only the low-mass region of the spectrometer. The 
result of this mass-ID method is shown in Fig. 1(a). Tritons populate the 
bump centered at about 2800 MeV/$c^2$, which was determined to contain 
about 3\% of other particles. 

Protons and negative pions emerging from secondary vertices are used 
to reconstruct $\Lambda(\rightarrow \pi^- p)$'s, which are finally 
identified by the value of the $\pi^- p$ invariant mass. Fig. 1(b) shows 
the invariant mass of $\pi^-p$ pairs detected in coincidence with 
tritons. A $\pi^-p$ reconstructed pair is assigned to belong to a 
$\Lambda$ decay when the $\pi^-p$ invariant mass is in the range 
1116$\pm$14 MeV/c$^2$. The distribution of background events is flat  
outside the peak thus allowing for estimation of the number of events 
inside the peak itself. Such background events constitute 0.63$\pm$0.67 
out of 40 $\Lambda$ events. All the events with a 
$\Lambda(\rightarrow\pi^-p$) and $t$ in the final channel, which fulfill 
the above ID requirements, are eventually reconstructed. 

A reconstructed $^6Li(K^-_{stop},\Lambda t)A'$ event is represented  
in Fig. 2, which also shows the front-view of the FINUDA layers. The 
right-hand side shows the central region of FINUDA (the DA$\Phi$NE 
beam-pipe, TOFINO, ISIM, the set of targets and OSIM), while the outer 
tracking region (LMDC's and ST) along with  TOFONE are shown on the 
left-hand side of the figure. An initial $\Phi(1020)$ meson decays to 
$\Phi\rightarrow K^+K^-$. The $K^-$ stops in the target ladder leading
%
%
\begin{figure}[b]
 \centering
  \includegraphics*[angle=0,width=0.9\textwidth]
    {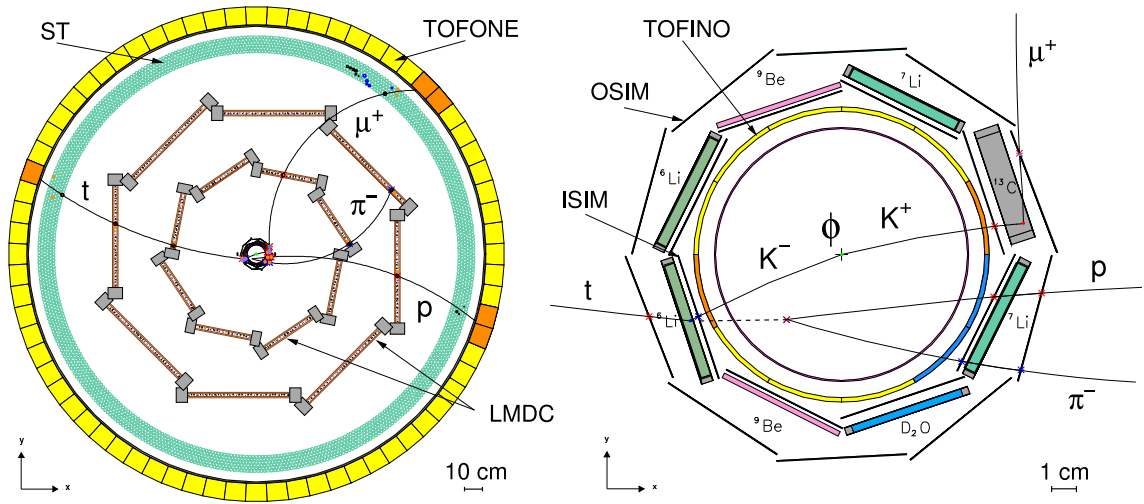}
       \caption{A $^{6}Li(K^-_{stop},\Lambda t)A'$ reconstructed event.
	 Particles are indicated in the figure, which also sketches the 
	 sensitive layers of FINUDA. The right-hand side represents the 
	 central region of the spectrometer, which closely surrounds 
	 the DA$\Phi$NE beam-pipe (the inner layer). The left-hand side 
	 depicts the outer tracker of FINUDA and TOFONE.}
\end{figure}
to a $\Lambda t$ pair; the $K^{+}$ decays to $K^+\rightarrow\mu^+\nu$. 
The (prompt) triton and $K^-$ tracks form a vertex, which identifies 
the stopping target, $^6Li$. A $\Lambda$ (dashed line) is emitted in a 
direction almost opposite to the triton direction, and travels for 
few centimeters before decaying to $\Lambda\rightarrow\pi^-p$. Some of 
the particles involved in the process, $\mu^+$, $p$ and $t$, impinge 
on  TOFONE. It can be noted that  the residual nucleus $A'$ is not 
reconstructed by FINUDA.

%
%
\begin{figure}[b]
 \centering
  \includegraphics*[angle=0.0,width=0.5\textwidth]
    {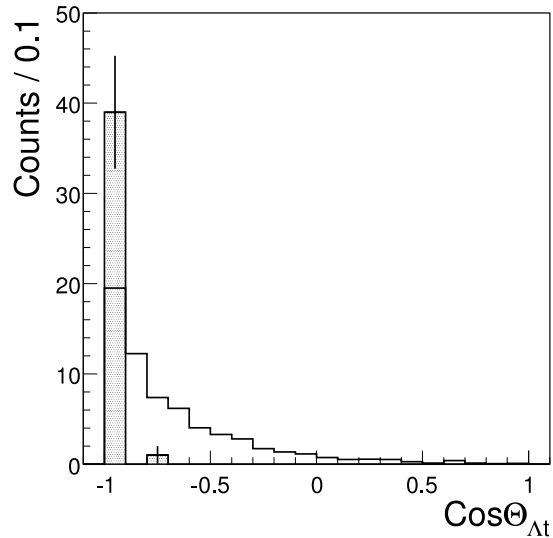}
    \caption{Opening angle distribution between  $\Lambda$ 
         and $t$ pairs (cos$\Theta_{\Lambda t}$). Filled histogram, 
	 experimental data; open histogram, phase space simulation for 
	 the $K^-_{stop} A\rightarrow\Lambda tN A'$ 
	 reaction.}
\end{figure} 
$\Lambda$-hyperons are detected in the momentum range from 140 MeV/$c$ 
(threshold) up to 800 MeV/$c$, whereas tritons in the $\Lambda t$ 
channel are analyzed starting from a momentum threshold of about 430 
MeV/$c$. $\Lambda$'s and $t$'s are measured with an average resolution 
$\Delta p/p<$2\% and $<$3\%, respectively. The given resolution for 
$p$'s (from $\Lambda$ decays) and $t$'s is the result of simulations,  
which are tuned to reproduce the momentum resolution $\Delta p/p<$0.7\%
of 236 MeV/$c$ muons from $K^+\rightarrow\mu^+\nu$. The $\Lambda$ angles 
were limited only by the geometric solid angle of FINUDA. For tritons, 
the forward requirement implies an azimuthal angle 
$0^\circ\le\Phi\le 180^\circ$, whereas the polar angle $\Theta$ depends 
only on the FINUDA angular acceptance, $45^\circ\le\Theta\le 135^\circ$ .

In this analysis, the reaction phase-space simulations are filtered 
through FINUDA. The experimental data are not corrected for the FINUDA 
acceptance; therefore, the results of simulations can be compared to the 
experimental points. The available data points (40) sum the contributions 
of six targets: 2$\times ^{6,7}$Li and 2$\times ^{9}$Be. Extensive 
simulations were performed. The simulated distributions were obtained by 
summing the outputs of each target and taking as weighting factors the 
number of $K^-_{stop}$'s in each target. Therefore, the notation $A$ in 
$K^-_{stop}A$ represents a generic nucleus.      
\section{The Results}
To understand the degree of correlation between $\Lambda$'s and $t$'s, 
the opening angle distribution cos$\Theta_{\Lambda t}$ is studied,
which is shown in Fig. 3 (filled histogram). The experimental data are 
sharply peaked at $-1$ thus indicating that correlated $\Lambda t$ pairs 
are mostly emitted in opposite directions. Since the data sum the 
contributions of all the targets, the cos$\Theta_{\Lambda t}$ narrowness 
indicates that the emission of such pairs is almost independent of the
nuclear target, and $\Lambda$'s and $t$'s are negligibly affected by 
Final State Interactions (FSI) in the different targets. In the 
following, when modeling the $K^-_{stop}A$ absorption, the phase space
simulated distributions are shaped to fit the narrow distribution
of the cos$\Theta_{\Lambda t}$ measured spectrum.

%
%
\begin{figure}[b]
 \centering
  \includegraphics*[angle=0,width=1.0\textwidth]
    {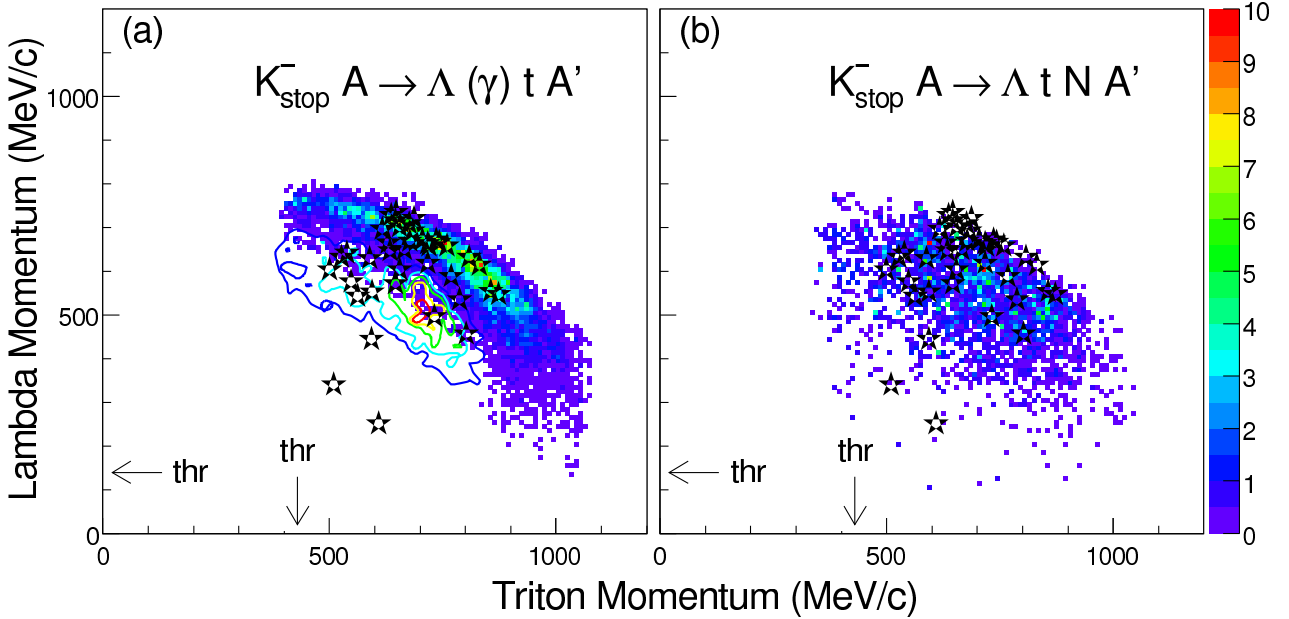}
       \caption{$p_{\Lambda}$ vs $p_t$ plots for the direct reactions 
       (a) $K^-_{stop}A\rightarrow\Lambda(\gamma)tA'$ and (b) 
       $K^-_{stop}A\rightarrow\Lambda t N A'$, where the notation $A'$ 
       represents a bound system of nucleons and $N$ a single nucleon. 
       The experimental data are represented by stars. In (a), the phase 
       space distribution for the $K^-_{stop}A\rightarrow\Lambda t A'$
       reaction is represented by a diffusion plot; instead, the contour 
       plot shows the $K^-_{stop}A\rightarrow\Lambda\gamma tA'$ phase space 
       where the $\Lambda\gamma$ pairs are the product of $\Sigma^0$ decays.}
\end{figure}      
The aim of our analysis is to examine to what extent the $\Lambda t$
experimental data overlap with the phase space of specific reaction 
channels. Fig. 4(a) shows the $p_{\Lambda}$ vs $p_t$ diffusion plot for 
the $K^-_{stop} A \rightarrow\Lambda(\gamma) t A'$ reactions, where the 
residual nucleus $A'$ is constrained to be bound. The experimental data 
are represented by stars. In this case, the data-points populate only 
part of the phase space region, which indicates that  the reaction favors 
final states formed by four or more bodies. With respect to this outcome, 
the phase space for the $K^-_{stop}A\rightarrow\Lambda t N A'$ four-body 
reaction is simulated and the result is shown in Fig. 4(b). Note that the 
data-points are spread over the high-intensity region of the reaction 
phase space. Such a strong correlation suggests that the data-points are 
consistent with a direct reaction mechanism with (at least) four bodies 
in the final state, $K^-_{stop} A \rightarrow \Lambda t N A'$. Another 
process capable of yielding final $\Lambda$ hyperons is the
$K^-_{stop} A\rightarrow\Sigma^0 t A'$ reaction followed by the $\Sigma^0$
decay, $\Sigma^0\rightarrow\Lambda\gamma$. The phase space of this process
is represented by  the colored contour lines in Fig. 4(a), which show a 
moderate overlap with the data-points. For this process, an additional $N$ 
in the final state lowers the contour lines since $N$ removes energy from 
the initial reaction, thus the overlap between the data-points and the 
phase-space is not augmented. For clarity of representation, the contour 
plot is not shown in Fig. 4(b). The  
$K^-_{stop}A\rightarrow\Sigma^0 tNA'$; $\Sigma^0\rightarrow\Lambda\gamma$  
process is therefore neglected in these analyzes.

Fig. 3 also shows the phase-space simulation of the $\Lambda t$ opening  
angle for the $K^-_{stop} A\rightarrow\Lambda tN A'$ reaction (open 
histogram). The histogram is arbitrarily scaled to the experimental 
data (filled histogram). The difference is remarkable: the phase-space 
events nearly span the full angular range with a monotonically 
decreasing trend, which is not capable of explaining the sharp 
back-to-back correlated behavior of the experimental data. 

%
%
\begin{figure}[c,b]
 \centering
  \includegraphics*[angle=0,width=1.0\textwidth]
    {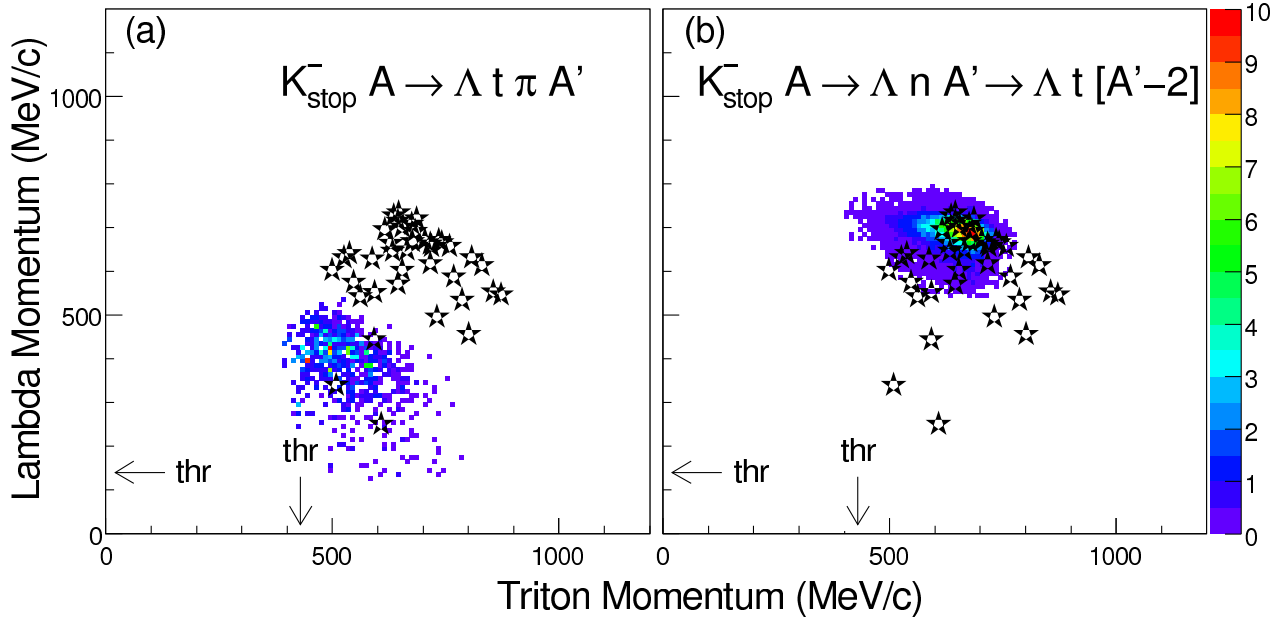}
       \caption{Diffusion plots of $p_{\Lambda}$ vs 
       $p_t$ for the multistep reactions (a) an on-shell pion is produced 
       in the final state $K^-_{stop} A\rightarrow\Lambda t\pi A'$ and (b) 
       a pick-up reaction $N A'\rightarrow t[A'-2]$ produces the  final 
       triton in $K^-_{stop} A\rightarrow\Lambda t [A'-2]$, where the 
       notation $[A'-2]$ represents a bound system of nucleons and $N$ a 
       single nucleon. The experimental data  are represented by  stars.}
\end{figure}       
Two-step processes may also provide strength to  $\Lambda t$ final 
states, although in general these processes are less probable than 
direct reactions since they require a further reaction to take place. 
Even so, such processes are modeled to test their degree of consistency 
with the measured data-points. At first, the elementary reaction 
$K^-_{stop}N\rightarrow\pi\Lambda$ is assumed to mediate the $\Lambda$ 
production in the reaction $K^-_{stop} A\rightarrow\Lambda t\pi A'$. The 
resulting reaction phase space is shown in Fig. 5(a) and it is compared 
to the data-points. There  is almost no overlap, thus disfavoring 
intermediate $\Lambda$'s being mediated by on-shell $\pi$'s. If these 
pions are absorbed while leaving the nucleus then they return their 
total energy to the reaction and the reaction phase space changes. In 
this case, if the absorbing cluster is an $\alpha$-substructure 
$\pi^- \alpha \rightarrow nt$ the momenta of the emerging tritons matches 
well the measured momenta. However, the $\Lambda$ hyperons coming from 
the $K^-_{stop}N\rightarrow\pi\Lambda$ reaction have $p_{\Lambda}<$300 
MeV/$c$, which is clearly lower than the measured $\Lambda$ momenta.
$\Lambda$ hyperons can also be produced by the two-body absorption 
reaction $K^-_{stop}np\rightarrow n \Lambda$. If the neutron picks 
up a deuteron  $n A'\rightarrow t[A'-2]$, or knocks 
out a triton $n A'\rightarrow n t[A'-3]$ then a final 
triton is available. This two-step process was modeled by requiring an 
intermediate pick-up (knock-out) reaction. Fig. 5(b) shows the results 
of the modeling when the $n A'\rightarrow t[A'-2]$ 
pick-up reaction takes place. The phase-space and data-points partially 
overlap. The overlap is less evident when modeling the knock-out reaction 
since a valuable amount of kinetic energy is taken away by the knocking-out 
neutron (figure not shown). The modeling of multistep reactions accounts 
for the Fermi motion of the initial $K^-_{stop}$-absorbing nucleon(s) and 
intermediate interacting clusters. For the $n+d\rightarrow t$ pick-up 
reaction, the approach of Ref. \cite{theor:planinic} was followed, which 
requires the relative momentum of $nd$ pairs to not exceed the triton 
Fermi momentum. 

The $K^-_{stop} A$ absorption rates leading to final $\Lambda t$ 
pairs are determined by accounting for the global acceptance of FINUDA. 
The values are listed in Tab. 1. These values can be compared with
the branching ratio reported in Ref. \cite{expt:roosen} for 
$K^-_{stop}$$^4He\rightarrow\Lambda t/ K^-_{stop}$$^4He\rightarrow all= 
0.0003\pm$0.0002, in which three events out of 3258 were found to be 
compatible with the $K^-_{stop}$$^4He\rightarrow\Lambda t$ reaction. 
However, the same three events could fit equally well the $\Lambda dn$ 
kinematics. For the sake of comparison, an absorption rate  of
$(4.4\pm 1.4)\times 10^{-3}/K^-_{stop}$ was determined  for the 
$K^-_{stop}$$^6Li\rightarrow\Lambda d A'$ reaction \cite{expt:FINUDA2}.
\begin{table}[b,c] 
\caption[Table]
{Absorption rates for $K^-_{stop}$ in $^6Li$, $^7Li$ and $^9Be$ to final 
$\Lambda t$ pairs. In the last column,  the number of reconstructed 
$\Lambda t$ pairs in each target is listed.}
\begin{center}
\vspace{0.3cm}
\begin{tabular}{ccc} \hline\hline
Nucleus &      Absorption rate           &  No. of events \vspace{-0.2cm}  \\ 
        &  [$\times 10^{-4}/K^-_{stop}$] &                                 \\ 
\hline
      $^6Li$         & $7.1\pm 3.4~(stat)~^{+1.2}_{-0.7}~(syst)$  &     7  \\
      $^7Li$         & $12.7\pm 3.7~(stat)~^{+2.1}_{-1.3}~(syst)$ &    15  \\
      $^9Be$         & $11.1\pm 2.9~(stat)~^{+1.8}_{-1.1}~(syst)$ &    18  \\
$^6Li + ^7Li + ^9Be$ & $10.1\pm 1.8~(stat)~^{+1.7}_{-1.0}~(syst)$  &   40  \\
\hline\hline
\end{tabular}
\end{center}
\end{table}
\section{Discussion and Conclusions}
This Letter presents novel results on $\Lambda t$ pairs arising from
the $K^-_{stop}A$ absorption reaction. The $\Lambda t$ 
signal is (nearly) free from background. The number of $\Lambda t$ 
pairs collected is low; nevertheless, a comparison of the $\Lambda t$ 
data with the phase space of several reactions makes it possible to 
draw several important conclusions. The phase space was constrained to 
reproduce the angular distribution of the detected  $\Lambda t$ pairs, 
that is, their strong back-to-back correlation. For the multistep 
reactions, the interacting nucleon and nucleon clusters were taken with 
their Fermi motion. The events resulting from the reaction modeling 
were reconstructed in FINUDA to account for the spectrometer acceptance. 
The comparison between data-points and simulations is based solely on 
the overlap of diffusion (contour) plots.

The phase space volume of the $K^-_{stop} A$ absorption reaction
yielding final pions barely overlaps the kinematic volume covered by the 
measured $\Lambda t$  pairs. This indicates that the $\Lambda t$ channel 
is primarily a non-pionic channel. 

The $p_\Lambda\; vs\; p_t$ correlation plot of the two-step process 
$K^-_{stop}(np)[A-2]\rightarrow n\Lambda[A-2]\rightarrow \Lambda t A'$ 
(Fig. 5(b)) overlaps some of the $p_\Lambda\;vs\;p_t$ measured data-points. 
This suggests that this multistep process may favor final $\Lambda t$ 
events. However, it requires (at least) two subsequent reactions to 
occur, which lessens the reaction probability to open the $\Lambda t$ 
channel with respect to the direct reaction  mechanism (Fig. 4).

The results obtained for the cos$\Theta_{\Lambda t}$ distribution 
indicate that the $\Lambda t$-pair production is strongly driven by 
reaction dynamics; in fact, phase-space simulations which are able to 
explain the $p_\Lambda\;vs\; p_t$ correlation plot are not capable of 
describing the cos$\Theta_{\Lambda t}$ behavior (Fig. 3). A reaction 
mechanism qualitatively capable of explaining the sharp peak and the 
number of nucleons involved requires that negative kaons are favorably 
absorbed by $\alpha$-like substructures of nuclei giving rise to the 
$K^-_{stop}\alpha\rightarrow\Lambda t$ reaction. 

The reactions $K^-_{stop} A\rightarrow\Lambda t (N)A'$ are largely 
capable of explaining the data-points (Fig. 4). They are partially 
explained by the $K^-_{stop}A\rightarrow\Lambda tA'$ reaction channel. 
By allowing a further nucleon in the final state, the reaction 
$K^-_{stop} A \rightarrow \Lambda t N A'$ complements the agreement. 
Such a nucleon might well be the consequence of a nuclear de-excitation 
similar to the process of negative pion absorption at rest in nuclei. 
In fact, the sharpness of the cos$\Theta_{\Lambda t}$ distribution 
indicates that $\Lambda$ and $t$ FSI are minimal. The mechanism for 
these reactions could be either a simple quasi-free transition, or a 
more composite process which embodies an intermediate bound system, 
$K^-_{stop}\alpha\rightarrow [K^- \alpha]\rightarrow\Lambda t$. 
Unfortunately, further analysis cannot be performed because of the 
poor statistics of the present data and, therefore, a specific reaction 
mechanism cannot be disentangled. Finally, the modeling of the process 
$K^-_{stop}A\rightarrow\Sigma^0 tA'$; $\Sigma^0\rightarrow\Lambda\gamma$ 
shows that phase-space and data-points moderately overlap. Accordingly, 
some of the $\Lambda t$ strength derives from it. 

The $K^-_{stop}A\rightarrow\Lambda p(d,t)NA'$ reactions cannot be 
explained  by looking only at a specific observable. In order to fully 
understand the reaction mechanism, including possible intermediate 
states (i.e., KNB states), many observables must be measured such as 
$\Lambda p(d,t)$ invariant  masses, angular correlations, $\Lambda$ 
and $p(d,t)$ momenta, absorption rates etc.. As well, theories must be 
able to account for most of the measured observables before giving a 
reliable view of the dynamics of the absorption process. In this regard, 
recent theoretical articles about the $K^-_{stop}A$ reaction meet  
these requirements \cite{theor:yamazaki,theor:oset}. 

The value determined for the $K^-_{stop}$$^{6}Li$ absorption rate is
$7.1\pm 3.4\times 10^{-4}/K^-_{stop}$. This value is compatible with 
the absorption rate previously published for $^4He$ 
$3\pm 2 \times 10^{-4}/K^-_{stop}$ \cite{expt:roosen}, which indicates 
that the measured process is likely due to direct multinucleon 
non-pionic K absorption.
\ack{
The present work was supported by the Istituto Nazionale di Fisica 
Nucleare (INFN) of Italy. One of us (GB) was also supported by NSERC
of Canada and through UVic by Dr. H. Jackh. The authors would  like to 
acknowledge the support received from the DA$\Phi$NE crew, which was 
highly appreciated. The collaboration from the INFN computing farm at 
Trieste helped to improve the quality of the analysis.
}
%
%
 
%
%
\end{document}